\documentclass[12pt,preprint]{aastex}

\slugcomment{To appear in Ap J. Letters}

\shorttitle{Molecular hydrogen as baryonic dark matter}
\shortauthors{Heithausen}

\begin{document}


\title{Molecular hydrogen as baryonic dark matter}


\author{Andreas Heithausen}

\affil{Radioastronomisches Institut, Universit\"at Bonn,
    Auf dem H\"ugel 71, 53121 Bonn, Germany}

\email{heith@astro.uni-bonn.de}

\begin{abstract}
High-angular resolution CO observations of small-area molecular
structures (SAMS) are presented.  The feature-less structures seen in
the single-dish measurements break up into several smaller clumps in
the interferometer map. At an adopted distance of 100pc their sizes
are of order a few hundred AU, some of which are still unresolved at
an angular resolution of about $3''$. The clumps have a fractal structure
with a fractal index between 1.7 and 2.0. Their kinetic temperature
is between 7~K and 18~K. Adopting standard conversion
factors masses are about 1/10 $M_{Jupiter}$\ for individual clumps and
densities are higher than 20000~cm$^{-3}$.  The clumps are highly
overpressured and it is unknown what creates or maintains such
structures. 
\end{abstract}



Received: 02 Mar 2004 - Accepted: 19 Mar 2004

\section{Introduction}

Measurements of the 3K microwave background in connection with
big-bang nucleosynthesis have impressively shown that most of the
matter in the universe is in some unknown non-baryonic form and only
16\% is baryonic in nature \citep{spergel2003}, of which only a small
part has been detected so far \citep{turner1999}. According to
gravitational lensing experiments less than 25\% of the unseen
baryonic matter in our Galaxy can be in form of massive compact halo
objects \citep{afonso2003}. An interesting alternative is molecular
hydrogen \citep{pfenniger1994,gerhard1996,walker1998}, because at most
temperatures in the interstellar medium it cannot be observed
directly, but only through secondary tracers such as carbon monoxide,
CO. \citet{pfenniger1994} have proposed that most of the dark matter
in the outskirts of our Milky Way could be in form of cold molecular
gas with a fractal structure. Basic building blocks in their model are
so-called clumpuscules with sizes of about 100AU and Jupiter mass 
($10^{-3}$~M$_\sun$).

Clumpuscules should form in the fragmentation process of an
interstellar cloud initiated by efficient cooling. Thus they should
exist throughout the Galaxy. In the outer region of the Galaxy they
are hard to detect, because due to their low metallicity, thus low CO
abundance, and low kinetic temperature they emit only very low
intensities, only slightly above the 3K cosmological background. In
the inner Galaxy conditions are more favourable to detect them in CO
lines, but here it is impossible to isolate individual clumpuscules
due to overcrowding in normal molecular gas.  The best location to detect
individual clumpuscules is at high galactic latitudes far away from
known molecular clouds.  

Good candidates for such structures are the small-area molecular
structures (SAMS) which were recently detected in a region with very
low extinction in the CO $J=2\to1$ and $1\to0$ lines
\citep{heithausen2002}.  In the low-angular resolution data they
appeared as unresolved clouds with full widths at half maximum below 1
arcmin. Adopting normal conditions one can show that such structures
cannot survive very long in the interstellar radiation field.  In this
{\it Letter} I will present high-angular resolution observations of
the SAMS, which show that they are composed of tiny molecular clumps,
which with respect to their structural properties resemble the
molecular clumpuscules proposed by \citet{pfenniger1994} as bayonic
dark matter candidate, however are warmer and probably less massive.

\section{Observations}

To further resolve the SAMSs high angular resolution data of one of
the structures described by \citet{heithausen2002} were obtained
simultaneously in the CO $J=2\to1$\ and $1\to0$\ line using the IRAM
Plateau de Bure interferometer near Grenoble in the French Alps.  SAMS2
was observed between January and October 2003 with the 6 15m
telescopes in the C and D configuration. The three-point mosaic covers
an area of $83'' \times 43''$\ with an angular resolution of $3.4''
\times 2.5''$\ at 115 GHz and $62'' \times 22''$\ with an angular
resolution of $2.5'' \times 2.1''$\ at 230~GHz. The final spectral
resolution is 0.1 km s$^{-1}$ at 115~GHz and 0.2 km s$^{-1}$ at 230
GHz, respectively.

\section{Results}

\begin{figure}
\epsscale{0.80}
\plotone{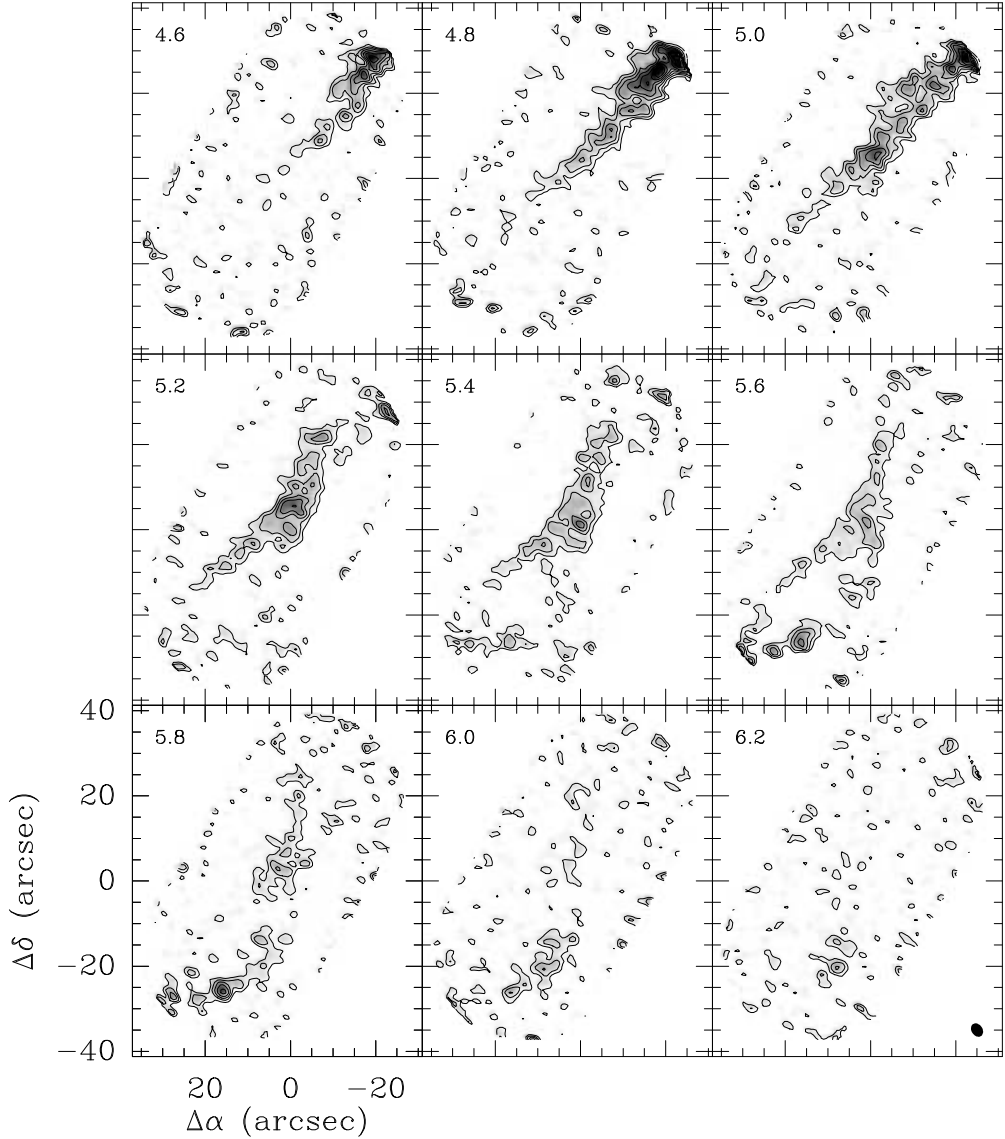}
\caption{
Distribution of the line intensity of individual velocity channels
with a width of 0.2 km s$^{-1}$\ in the CO $J=1\to0$ transition; the
center velocity of each submap is indicated in the upper left corner
of each box. Contours are every 0.64 K km s$^{-1}$ starting at 0.64 K
km s$^{-1}$ ($3\sigma$); the noise level increases slightly towards
the borders of the map.  Note that the most intense clump at the upper
border of the maps at velocities between 4.6 and 5.0 km s$^{-1}$\ is
at the edge of the primary beam of a single telescope of the
interferometer which might artifically sharpen its outer edge.
Offsets are relative to $\alpha_{J2000}=9^h56^m58^s$, $\delta_{J2000}=
69^\circ19'35''$. The angular resolution of the maps is indicated by a
black ellipse in the lower right corner of the lower right box. It
corresponds to 340AU by 250AU at the adopted distance of 100pc.  }
\label{chanmaps}
\end{figure}

The resulting channel maps of SAMS2 in the CO ($1\to0$) transition are
displayed in Figure \ref{chanmaps}. The 2 featureless structures seen
in the single-dish map split up in the interferometer map into a
smaller filamentary structure composed out of even smaller
clumps. Some of the clumps are unresolved even in the high-angular
resolution map. In order to derive physical parameters for the most
intense structures  the data set was decomposed into gaussian shaped
clumps using the GAUSSCLUMP algorithm developed by
\citet{stutzki1990}. Table \ref{table1} lists the fitted values of the
relative positions, $\Delta x$ and $\Delta y$, radii, amplitudes,
velocities and line widths. The radii listed are geometric means of
the full-widths at half maximum of a 2-dimensional gaussia, fitted to
the intensity distribution. They have been deconvolved for the
interferometer resolution; Peak temperature and density have been
corrected for this effect.  Mass and average density are derived
adopting a distance of 100pc and assuming a CO to H$_2$ conversion
factor of $X=1.5\times 10^{20}$ cm$^{-2}$ (K km s$^{-1})^{-1}$,
i.e. the galactic value \citep{hunter1997}. Note that the
uncertainties in these assumptions allow only an order of magnitude
estimate.

The total mass of all structures in the map correspond to 1.8
$M_{Jupiter}$, consistent with the value from the single dish data
derived using the same assumptions. The fact that the total flux of
the single dish observations is completely recovered with the interferometer
implies that there is not much structures on larger scales because
with the setup chosen the interferometer is only sensitive to
structures below an angular resolution of $26''$ or to linear structures
of below 2600AU at the adopted distance. The most intense clumps
listed in Table \ref{table1} account for 0.6 $M_{Jupiter}$\ with
individual masses of less than 1/10 $M_{Jupiter}$ The fractal index,
$D$, of the structure is thus
\begin{equation}
D={{\mathrm{log}}(M/M_0)\over {\mathrm{log}}(r/r_0)}=1.7-2.0
\end{equation}
similar to that found for galactic molecular clouds
\citep{falgarone1991}; here $r$ and $r_0$ correspond to the different
angular scales and $M$ and $M_0$ are the masses of identifiable
objects at those scales. Note that because the fractal index depends only on 
the ratios of masses and sizes it is independent on actual distance and
the way the mass is determined.

The line width of the individual structures is about 0.4 km s$^{-1}$,
a factor 2 lower than that seen in the lower angular resolution
map. This difference results from a monotonous velocity gradient of 1
km s$^{-1}$ over a distance of only 6000~AU. This enormous gradient is ten
times higher than that observed for most other molecular clouds and
their embedded cores \citep{goodman1993,grossmann1992}. It
implies that the structures are dynamic objects far from static
equilibrium. This aspect was predicted for clumpuscules by
\citet{pfenniger1994}, because due to their inhomogeneous fractal
structure clumpuscules should collide frequently which prevents them
from collapsing to Jupiter like objects.

All the structures seen in the CO ($1\to0$) transition are also
detected in the CO ($2\to1$) transition, however due to athmospheric
conditions with a lower signal-to-noise ratio. The ratio of the
integrated line intensities of the two transitions is $R(2\to1/1\to0)
= 0.7\pm0.2$. Under the assumption of optically thick lines the
excitation temperature is $T_{ex}=7.0-18.2$~K using the values from
Table \ref{table1}. It provides a good estimate for the kinetic
temperature. The linewidth of the individual clumps
(s. Tab. \ref{table1}) is much larger than the thermal linewidth
indicating the possibility of further substructure as also indicated
in the maps (s. Fig. \ref{chanmaps}). The clumps are highly
overpressured; with $P/k=(0.1-1.5)\times10^6$~K cm$^{-3}$ they exceed
the average interstellar pressure \citep{jenkins2001} by at least 1 to
2 orders of magnitude. This raises the question how such structures
can form or survive if not stabilized by their own gravity.

\section{Discussion}

In recent years molecular hydrogen has been detected in the diffuse
interstellar medium directly via absorption-line measurements towards
many distant quasars \citep{jenkins2003,richter2003}. 
Such observations, which trace mainly warm
gas with low column densities, show that H$_2$ is wide spread in the
Galaxy even outside star-forming regions. They however provide only
little information on the spatial structure of the clouds. The
emission-line observations presented here for the first time disclose
that molecular clouds in the diffuse interstellar medium are fractally
structured. The building blocks of such clouds are low mass clumps or
clumpuscules with sizes of order hundred AU.  Due to the high
uncertainty in mass and density it is impossible to judge whether
these clumps are stabilized by their own gravity. The mass
determination is based on conditions as for an average molecular cloud
in the galactic plane, which may not hold at such small
scales. Further excitation studies of the clumps are required to solve
this issue.

It is interesting to note that the clumps have the same radial
velocity as the surrounding atomic gas \citep{heithausen2002} and are
within the range of velocities for the molecular gas in the Ursa Major
cirrus clouds \citep{devries1987} which are a few degrees away from
the clumps. This means that tiny molecular clumps are possibly a
natural constituent of the diffuse interstellar medium, however not
recognized as such so far. They can form or survive even in regions
with low column densities. From $^{12}$CO and $^{13}$CO 
 observations of molecular gas in translucent clouds the
existence of small scale structure down to a few hundred AU has been
inferred \citep{falgarone1998}; the observation presented here
indicate that such structures are possibly intrinsically linked to the
formation process of molecular clouds.

Whether or not fractally structured clouds similar to the ones
described here could account for all the missing baryonic dark matter
in our Galaxy is hard to determine. 
To be consistent with the $\gamma-$ray background 
denser and more massive clumps than described in this {\it Letter}
are predicted by models, which explain 
the missing baryonic mass with molecular clumpuscules
\citep{depaolis1999,kalberla1999}. 
Due to their small sizes and their
narrow line widths both, the hypothetic dark matter clumpuscules and
the clumps described here, are notoriously hard to detect unless they
form small clusters and thus fill at least some part of the beam of a
larger telescope. With small telescopes used to survey the Galaxy they
are undetectable due to the low beam filling; high angular resolution
and high sensitivity observations are needed for detection. Such
clouds thus provide an ideal means to hide matter from the observer.

\acknowledgments

I thank Arancha Castro-Carrizo and Roberto Neri for their help during
the reduction of the PdB data and Philipp Richter and Peter Kalberla
for critically reading the manuscript. This letter is based on
observations with the IRAM Plateau der Bure Interferometer. IRAM is
supported by the INSU/CNRS (France), MPG (Germany) and IGN (Spain).

%\clearpage

%% copy editing.  Individual instruments can be provided in parentheses,
%% after the keyword, but they will not be verified.

Facilities: \facility{PdB}.

\clearpage

\begin{deluxetable}{crr crc ccc}
%\tabletypesize{\scriptsize}
%\rotate
\tablecaption{Parameters for the most intense clumps.}
%\tablewidth{0pt}
\tablehead{
\colhead{\#} & \colhead{$\Delta x$} & \colhead{$\Delta y$} & \colhead{Radius} & 
\colhead{$T_{mb}$} &
\colhead{$v_{LSR}$} & \colhead{$\Delta v$} & \colhead{Mass} &
\colhead{Density} \\
& \colhead{($''$)}	&\colhead{($''$)} & \colhead{(AU)}	&\colhead{(K)}	&\colhead{(km s$^{-1}$)} 
&\colhead{(km s$^{-1}$)} 
&\colhead{($M_{Jupiter}$)}&\colhead{(cm$^{-3})$}
}
\startdata
 1  &	-22.2   &	28.4 &  290  &	10.5 &  4.8 &  	0.46 &   0.073 &   66000\\
 2  &	-18.6   &	25.0 &  410  & 	7.6  & 	4.8 &  	0.39 &   0.089 &   39000\\
 3  &	-12.8   &	21.3 &  630  & 	4.0  & 	4.8 &  	0.42 &   0.118 &   17700\\
 4  & 	-1.8    &	6.2  & 	820  & 	3.8  & 	5.0 &  	0.52 &   0.240 &   17400\\
 5  &  	0.0    	&	1.2  & 	410  & 	4.1  & 	5.4 &  	0.36 &   0.045 &   19500\\
 6   &	15.9  	&	-26.1 &  160 &  14.8 &  5.8 &  	0.49 &   0.033 &   84000\\
\enddata

\tablecomments{Adopted distance 100pc; adopted $X=1.5\times10^{20}$ cm$^{-2}$ 
(K km s$^{-1})^{-1}$.  
Offsets are relative to $\alpha_{J2000}=9^h56^m58^s$, $\delta_{J2000}= 
69^\circ19'35''$. 
The radius has been deconvolved from the instrumental resolution; 
temperature and density are corrected for this effect.}

\label{table1}
\end{deluxetable}

\clearpage

%% This figure uses \includegraphics to scale and rotate the still frame

\begin{thebibliography}{}

\bibitem[Afonso(2003)]{afonso2003}
Afonso, C., Albert, J.N., Andersen, J., et al., 2003, \aap\ 400, 951.

\bibitem[De Paolis et al.(1999)]{depaolis1999}
de Paolis, F., Ingrosso, G. Jetzer, P., Roncadelli, M., 1999, 
\apj\ 510, L103.

\bibitem[De Vries et al.(1987)]{devries1987}
de Vries, H.W., Heithausen, A., Thaddeus, P., 1987,
\apj\ 319, 723.

\bibitem[Falgarone et al.(1991)]{falgarone1991}
Falgarone, E., Phillips, T.G., Walker, C.K., 1991, \apj\  378, 186.

\bibitem[Falgarone et al.(1998)]{falgarone1998}
Falgarone, E., Panis, J.F., Heithausen, A., et al. 1998, \aap\ 331, 669.

\bibitem[Gerhard \& Silk(1996)]{gerhard1996}
Gerhard, O., Silk J., 1996,  \apj\  472, 34.

\bibitem[Goodman et al.(1993)]{goodman1993}
Goodman, A.A. Benson, P.J., Fuller, G.A., Myers, P.C., 1993, \apj\  406, 528.

\bibitem[Gro{\ss}mann \& Heithausen(1992)]{grossmann1992}
Gro{\ss}mann, V., Heithausen A., 1992, \aap\ 264, 195.

\bibitem[Heithausen(2002)]{heithausen2002}
Heithausen, A., 2002, \aap\ 393, L41.

\bibitem[Hunter et al.(1997)]{hunter1997}
Hunter, S.D., Bertsch, D.L., Catelli, J.R., et al., 1997,  \apj\  481, 205.

\bibitem[Jenkins et al.(2001)]{jenkins2001}
Jenkins, E.B., Tripp, T.M., 2001, \apj\  Suppl. 137, 297.

\bibitem[Jenkins et al.(2003)]{jenkins2003}
Jenkins, E.B., Bowen, D.V., Tripp, T.M., et al., 2003, \aj\ 125, 2824. 

\bibitem[Kalberla et al.(1999)]{kalberla1999}
Kalberla, P., Shchekinov, Y.A., Dettmar, R.-J., 1999,
\aap\ 350, L9.

\bibitem[Pfenniger \& Combes(1994)]{pfenniger1994}
Pfenniger, D. \& Combes, F., 1994,  \aap\ 285, 94.

\bibitem[Richter et al.(2003)]{richter2003}
Richter, P., Wakker, B.P., Savage, B.D., Sembach, K.R., 2003,  \apj\  586, 230.

\bibitem[Spergel et al.(2003)]{spergel2003}
Spergel, D.N., Verde, L., Peiris, H.V., et al., 2003,  \apj\  Suppl. 148, 175.

\bibitem[Stutzki \& G\"usten(1990)]{stutzki1990}
Stutzki, J. \& G\"usten, R., 1990,  \apj\  356, 513.

\bibitem[Turner(1999)]{turner1999}
Turner, M.S., 1999, \pasp\ 111, 264.

\bibitem[Walker \& Wardle(1998)]{walker1998}
Walker, M., Wardle, M., 1998, \apj\  498, L125.

\end{thebibliography}
\end{document}